# A Secure File Sharing System Based on IPFS and Blockchain

Hsiao-Shan Huang
Department of Electronics Engineering, National Chiao-Tung University
Telecommunication Laboratories, Chunghwa Telecom Co., Ltd.
Hsinchu, Taiwan
phm@cht.com.tw

Tian-Sheuan Chang
Department of Electronics Engineering, National Chiao Tung University
Hsinchu, Taiwan
tschang@mail.nctu.edu.tw

Jhih-Yi Wu
Telecommunication Laboratories, Chunghwa Telecom Co., Ltd.
Taoyuan, Taiwan
ian_wu@cht.com.tw

## ABSTRACT
There is a great interest in many approaches towards blockchain in providing a solution to record transactions in a decentralized way. However, there are some limitations when storing large files or documents on the blockchain. In order to meet the requirements of storing relatively large data, a decentralized storage medium is produced. IPFS is a distributed file system which is content-addressable. It works very similar to the blockchain network. There are some attempts which take advantage of the blockchain concept and IPFS to design new approaches. Unfortunately, there are some inefficiencies in sharing data using the combination of IPFS and blockchain. In this paper, we proposed a secure file sharing system that brings a distributed access control and group key management by the adoption of the IPFS proxy. The IPFS proxy which plays an important role in the design is adopted to take responsibility for the control policies. The combination of the IPFS server and the blockchain network with the adoption of the IPFS proxy make a secure file sharing system which the members on the system can create new groups or join different groups by their own choice. Although there is no access control mechanism in IPFS server and blockchain network, the secure file sharing system manages the access control policies. The members access files only belong to the group they authorized.

## 1. INTRODUCTION
In recent years, blockchain has become one of the most popular innovations. It first came to light in 2008 and is generally known as Bitcoin[1]. Blockchain brings a new perspective on making a trusted transaction between two entities without a third party being participated in. The main advantage of blockchain is to reduce risk of tampering due to its immutable nature. It also has several other advantages. Decentralization means no single entity controls the system. Trackability and incorruptibility indicate the transaction results are linked together by utilizing cryptographic mechanism so that the results on the chain are undeniable.

In addition to blockchain, Interplanetary File System(IPFS)[2] is another attractive design which is a peer-to-peer distributed file system. It combines several past successful systems and provides a content-addressed block storage model. The propose of IPFS is to improve HTTP, which is one of the most dominant file distributed system nowadays.

There are several attempts to take advantage of the blockchain concept and IPFS to design new approaches. [3] proposed a blockchain-based storage system for sharing IoT data. It adopts a re-encryption token[4] to manage the keys so that the data encryption key is updated to avoid further access to the system when an authorized user is revoked. However, the key encryption and update processes are all handled by the data owner, which significantly increases the computation overheads of the data owner. [5] proposed a scalable and trustworthy file sharing system. It uses another re-encryption technique[6] and provides a scalable key management scheme to support the accessibility of multiple users. The encryption key is stored in a distributed way which can resist collusion attacks by revoked users. Since the one sub key of the encryption key is sent to the authorized user, the key must be managed by the user which increases the security risks.

The above schemes have some insufficiencies and limitations. For improving the computation overheads and security risks, this paper introduces a secure file sharing system based on IPFS and blockchain. In which, we adopt a IPFS proxy for the distributed access control policies and group key management. The files encrypted by the IPFS proxy can be stored on the IPFS server by taking advantage of the decentralized file distributing system of IPFS servers. Besides, we use blockchain as a trackable server to record the file uploading information in the network and thus avoid information corruption. The combination of the IPFS server and the blockchain network with the adoption of the IPFS proxy make a secure file sharing system so that the members on the system can create new groups or join different groups by their own choices and only group members can access the authorized files.

The rest of the paper is organized as follows. In Section II, the background of blockchain and IPFS technologies are discussed. Section III presents the proposed secure file sharing system and describes how to combine the blockchain and IPFS to take advantage of their benefits. Finally, the conclusion and future work are presented in Section IV.

## 2. BACKGROUND
In this section, some technologies related to the proposed scheme are described below.

### 2.1 Blockchain
A blockchain is regarded as a distributed ledger that the records on the chain are maintained by the participating peers. The transaction results are validated by the peers and are linked together by utilizing cryptographic mechanism.

The first example of a blockchain implementation is Bitcoin[1]. Bitcoin is a digital cryptocurrency. The main idea of Bitcoin is to offer a decentralized banking system. Users exchange their assets and trust the transaction results which are recorded on the distributed ledger. The results are trusted since they are validated by the participants. After the validation, the transactions are transmitted to the blockchain which cannot be tampered with. The participants, known as the miners in the blockchain network, are rewarded with cryptocurrency for their computational work. This is so-call an incentive mechanism to validate transactions and thus make a consensus in the network. Since the miners are all anonymous peers, any single miner cannot be trusted to prevent the transactions results are dominated by the minority. Therefore, a consensus mechanism called proof-of-work is introduced to reach an agreement in the blockchain network. Generally, a new block contains a set of new transactions, information required to keep track of the history of transactions and the result of a mathematical puzzle. The miners do the computational work to find the answer of the puzzle. After solving the problem, the answer is recorded in the block and the new block is considered as a valid block. Anyone in the blockchain network can answer the puzzle by taking the computational work to earn transaction fees. This is described as the proof-of-work mechanism.

As described above, Bitcoin is an open blockchain which means it is permission-less. Any user can easily participate in the network to create transactions or take part in the consensus process. They can be a miner or a client. Ethereum[7] is also an open blockchain. Another rather different chain is consortium blockchain. It is controlled by a set of authorized nodes. These nodes dominate the validation process. Since the nodes are authorized in the network, it is called permissioned blockchain. Hyperledger[8] is the most popular case. It is different from permission-less blockchain since a trusted central party is required to validate the nodes who want to join the network.

As mentioned before, a blockchain is a distributed ledger and the records on the chain contains a set of new transactions, information required to keep track of the history of transactions etc. However, there are some limitations when storing large files or documents on the blockchain. It is essentially an expensive database for storing large amounts of data since the size of data in a block is limited. In order to meet the requirements of storing relatively large data, a decentralized storage medium is introduced such as IPFS [2], Filecoin[9], and Swarm[10]. Among them, IPFS is a distributed file system which is content-addressable. It works very similar to the blockchain network. Therefore, it is introduced below since the proposed method is related to IPFS.

### 2.2 IPFS
IPFS is a decentralized file distributing system. It is based on a peer-to-peer protocol. Each stored file is allocated a unique hash according to the content of the file. The designated hash is regarded as the corresponding file address when user requests the file. The deduplication mechanism is adopted to implement the decentralized concept in IPFS. The stored files are permanent in the system as long as there are some duplicate copies of the files.

There are some main properties which IPFS combines. First, Distributed Hash Tables(DHT) in IPFS is a data structure that the data is efficiently coordinated in order to quickly search between nodes. With DHT, the data can be shared without central coordination due to advantages of decentralization and scalability. Second, a peer-to-peer file sharing system called BitSwap is implemented in IPFS. It is similar to BitTorrent but different. BitSwap operates as a marketplace where the blocks can be acquired by nodes. The blocks that are part of a file may come from unrelated files in the filesystem. A Merkle directed acyclic graph(DAG) is another important feature of IPFS. It links between data blocks which are cryptographic hashes of the target. The Merkle DAG ensures the identification, and tamper resistance of the stored data. Along with Merkle DAG, Version Control Systems(VCS) is another feature of IPFS that allows users to easily management files. It allows users to edit different versions of a file and distribute multiple versions efficiently. Finally, the Self-Certifying File System (SFS) is the last concept of IPFS. It creates the Interplanetary Name Space (IPNS) which is an SFS to allows users verifying objects published on the network. The objects can be signed by the publishers with their private keys and therefore can be verified using the publisher's public key.

## 3. PROPOSED SCHEME
We propose a secure file sharing system based on permission-less blockchain and IPFS. The permission-less blockchain allows any user who wants to participate in the blockchain network. Since there is no trusted central party to validate users, the file sharing mechanism in the system may have permissions issue. The proposed design brings a distributed access control and group key management by the adoption of the IPFS proxy.

### 3.1 Motivation
Blockchain is regarded as a distributed ledger which provides a solution to record transactions in a decentralized way. However, nodes in the blockchain network have memory limitation and only the transaction records rather than a complete file can be stored on the blockchain. Therefore, IPFS is introduced to meet the requirements of storing relatively large data. On the other hand, computation overheads and security risks are not considered in the IPFS and blockchain. Consequently, we have gained motivation to improve the above issues. A secure file sharing scheme is proposed in this paper to manage the access control policies by utilizing the IPFS proxy.

### 3.2 System Model
Figure 1 shows the system model that consists of the following entities: owner, user, IPFS proxy, IPFS server and blockchain.

*3.2.1 Owner*
The owner is the one who creates a new group in the system. It is the first member in the group and has the rights to share files in a group. It also manages the access policies.

*3.2.2 User*
When the user becomes a group member, it has the rights to access the files which shared by the members in the group.

*3.2.3 IPFS Proxy*
A group key pair is generated in the IPFS proxy when a new group is created. Then the group key encrypts the file in order to preserve privacy of the file uploaded by the group member. The

IPFS proxy controls the mapping table that are key lists of the members in the group. When authorized member makes a request for a file, it encrypts the group key with the member's public key and sends the encrypted key to the authorized member.

### 3.2.4 IPFS Server

After encrypted by IPFS proxy, the encrypted files are stored in the IPFS server. When authorized member requests a file, the IPFS server returns the encrypted file to the member.

### 3.2.5 Blockchain

According to the characteristic of blockchain that is considered incorruptible, the file uploading information can be recorded on the blockchain. Along with the identification and hash value, the blockchain is regarded as a trackable server so that the ownership and integrity of a file is undeniable.

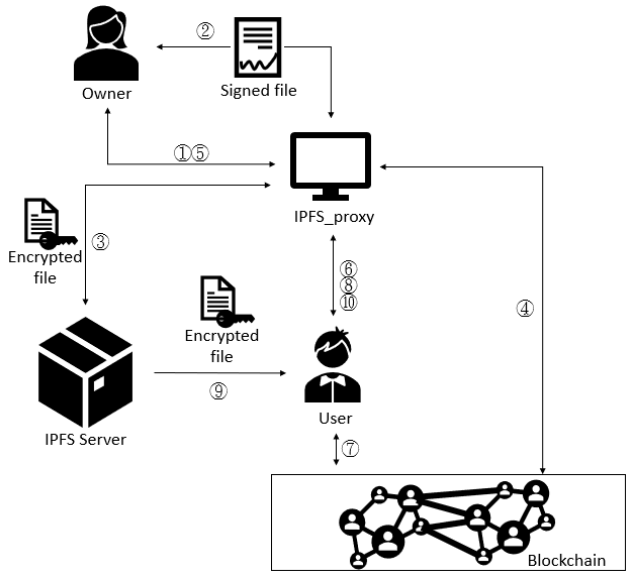

**Figure 1. Overview of the proposed secure file sharing system**

## 3.3 Work Flow

The following is a detailed description on the work flow of the proposed scheme.

In the first step, the owner who wants to create a new group in the secure file sharing system needs to register at the IPFS proxy. The owner uploads the public key generated by himself along with the user identification *user_id*. After the registration, the IPFS proxy returns a group identification *group_id* to the owner. The IPFS proxy owns a mapping table of the *group_id* mapping that are key lists generated by whom registered to the system.

In the step 2, the registered owner signs his file that is ready to be uploaded. The IPFS proxy verifies the correctness of the signature. After checking the signature properly, the IPFS proxy hashes the file and generates a *file_hash* to verify the integrity of the file. It also encrypts the file with a group key generated by the proxy. The group key is stored in the proxy and the encrypted file is uploaded to the IPFS Server. The IPFS Server returns a *IPFS_hash* after storing the encrypted file, as shown in the Step 3. Upon receiving the *IPFS_hash*, the IPFS proxy stores the combined value of *group_id*, *user_id* and the *file_qhash* on the blockchain, as shown in the Step 4. The blockchain then broadcasts a transaction and returns a transaction identification *trans_id* to the IPFS proxy. After returning *trans_id* from the IPFS proxy to the owner as shown in the Step 5, the uploading process is completed. The sequence diagram for file uploading process is shown in Figure 2.

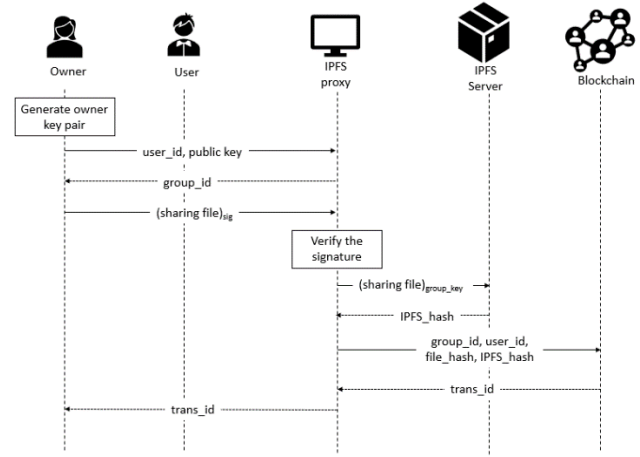

**Figure 2. Sequence diagram for file uploading process**

When a user wants to access the files belonging to a *group_id*, the user first generates a key pair and upload the public key along with *user_id* and the designated *group_id* to the IPFS proxy, as shown in Step 6. A new user needs to receive permission by the group owner. If the user is authorized to access files in the group, the IPFS proxy updates the user's public key to the mapping table and the user registration process is completed.

In the Step 7, the authorized user queries transaction data on the blockchain according to *trans_id*. After successful receive of the designated transaction data which including *group_id*, *user_id*, the *file_hash* and the *IPFS_hash*, the user signs the request to the IPFS proxy with *group_id*, *user_id* and *IPFS_hash*, as shown in the Step 8. The IPFS proxy verifies the identity of the request, and queries the IPFS server according to the *IPFS_hash*. Finally, the IPFS server send the designated encrypted file to the user, as shown in the Step 9.

Now the user cannot access the encrypted file since the file is encrypted by the group key generated by the IPFS proxy. We propose a key wrapping technique which encrypts one key using another key. To decrypt the file that the user received, the IPFS proxy sends a key wrapping key to the user. The key wrapping key encrypts the group key by the user's public key. The user then decrypts the key wrapping key and receives the group key. After that, the encrypted file can be decrypted by using the group key, as shown in the Step 10. Finally, the user hashes the file and compares the result with the *file_hash* downloaded from the blockchain to verify the integrity of the file. After the comparison, the sharing process is completed. The sequence diagram for file downloading process is shown in Figure 3.

## 3.4 Access Control

The proposed IPFS proxy plays an important role in the system. It provides a registration and validation platform to whom wants to attach to the secure file sharing system.

After the registration, the user's public key is recorded in a mapping table stored in the IPFS proxy. The IPFS proxy first generates a new group key when a new owner requests for registration in the system. The group key is a symmetric key, e.g., AES-CBC, since a symmetric cryptosystem is faster. The group key is stored in the mapping table along with the owner's public key, as shown in Figure 4. When another user wants to access to the same group, the user's public key is attached to the same mapping table after the validation is completed. The different groups have their own mapping tables that become the access control polices of the IPFS proxy.

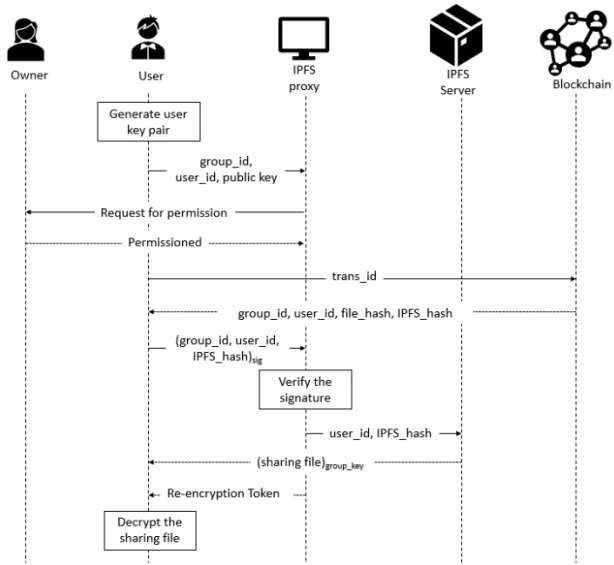

Figure 3. Sequence diagram for file downloading process

## 3.5 Revocation

All the sharing files in the same group are encrypted by the same group key. If there is someone in a group being revoked, then the IPFS proxy generates a new group key to replace the old one. All the encrypted files in the IPFS server that belong to the group require to re-encrypt by the new group key. The *IPFS_hash* of the new encrypted files are stored to the blockchain as a new transaction. All the users excluding the revoked user will receive the new group key encrypted by the public key of the authorized users. To prevent the revoked user to access any future data, the new sharing files are encrypted by the new group key. Given *n* users in a group, the revocation process requires a communication overhead in the order of *O(n)*.

## 3.6 Group Management

Any user in the system can join more than one group. For example, Alice, Bob and Carol belong to group 1, and Bob and Charlie are in group 2, as shown in Figure. 4. Bob belongs to both group 1 and group 2. When Bob wants to access files in the group 1, he requests the IPFS proxy for the group 1 key encrypted by his public key. Then Bob can decrypt the file downloaded from the IPFS server by group 1 key. In the same case, Bob can access the files belonging to group 2 by requesting the IPFS proxy for the group 2 key. Bob is the only person who can both access the files in group 1 and group 2.

Since Alice and Bob are in the same group, Alice only has the right to access the files in the group 1. In a similar way, Charlie only has permission to access the files in group 2 while Bob in the same group is authorized to access both groups. The group management classifies the members in the system according to group it belongs, regardless of how many groups a member belongs to.

## 3.7 Security

In our system, the IPFS proxy is trusted. The owners and users who want to join in the secure file sharing system are regarded as trusted after completing the registration process. However, the IPFS server and Blockchain are unreliable since their data is accessible to anyone on the internet. Despite the unauthorized users are available to access files from the IPFS server and Blockchain, the encrypted files downloaded from IPFS server cannot be decrypted without the group key. The IPFS proxy manages the access control polices to ensure the group security in the secure file sharing system.

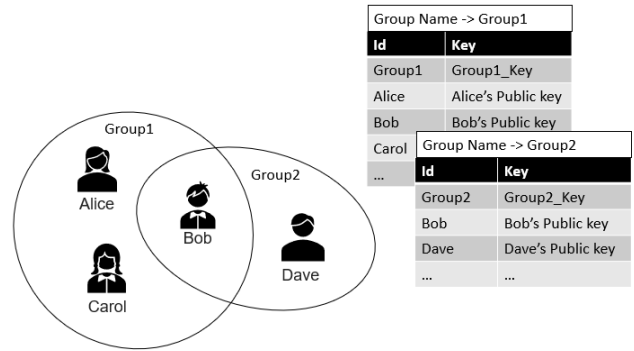

Figure 4. Group key management and the mapping table in the IPFS proxy

## 4. CONCLUSION AND FUTURE WORK

In this paper, we present a secure file sharing system based on permission-less blockchain and IPFS. The IPFS proxy is adopted to bring a distributed access control and group key management. The combination of the IPFS server and the blockchain network with the adoption of the IPFS proxy make a secure file sharing system. In which the members on the system can create new groups or join different groups by their own choice. Although there is no access control mechanism in the IPFS server and blockchain network, the secure file sharing system manages the access control policies. The members access files only belonging to the group they are authorized. We are currently in the process of finishing a implementation of our design. As for the future work, we will apply the design and show the detailed performance to prove our work.

## 5. REFERENCES

[1] Nakamoto, Satoshi, and A. Bitcoin. "A peer-to-peer electronic cash system." Bitcoin.–URL: https://bitcoin. org/bitcoin. pdf (2008).

[2] Benet, Juan. "Ipfs-content addressed, versioned, p2p file system." arXiv preprint arXiv:1407.3561 (2014).

[3] Shafagh, Hossein, et al. "Towards blockchain-based auditable storage and sharing of IoT data." Proceedings of the 2017 on Cloud Computing Security Workshop. 2017.

[4] Ateniese, Giuseppe, et al. "Improved proxy re-encryption schemes with applications to secure distributed storage." ACM Transactions on Information and System Security (TISSEC) 9.1 (2006): 1-30.

[5] Cui, Shujie, Muhammad Rizwan Asghar, and Giovanni Russello. "Towards blockchain-based scalable and trustworthy file sharing." 2018 27th International Conference

on Computer Communication and Networks (ICCCN). IEEE, 2018.

[6] Dong, Changyu, Giovanni Russello, and Naranker Dulay. "Shared and searchable encrypted data for untrusted servers." Journal of Computer Security 19.3 (2011): 367-397.

[7] Ethereum White-Paper. Online: https://github.com/ethereum/wiki/wiki/ White-Paper, 2019.

[8] Cachin, Christian. "Architecture of the hyperledger blockchain fabric." Workshop on distributed cryptocurrencies and consensus ledgers. Vol. 310. 2016.

[9] Protocol Labs. Filecoin: A Decentralized Storage Network. Online: https://filecoin.io/filecoin.pdf, 2017.

[10] Hartman, John H., Ian Murdock, and Tammo Spalink. "The Swarm scalable storage system." Proceedings. 19th IEEE International Conference on Distributed Computing Systems (Cat. No. 99CB37003). IEEE, 1999.